\documentclass[reprint,amsmath,amssymb,superscriptaddress]{revtex4-1}

\usepackage{graphicx}
\usepackage{dcolumn}
\usepackage{bm}
\usepackage{epstopdf}
\usepackage{xr}
\usepackage{xcolor}

\begin{document}

\title{On the relationship between orbital moment anisotropy, magnetocrystalline anisotropy, and Dzyaloshinskii-Moriya interaction in W/Co/Pt trilayers}

\author{Zhendong Chi}
\thanks{Corresponding author: zhdchi93@gmail.com; Present affiliation: Mitsubishi Electric Corporation.}
\affiliation{Department of Physics, The University of Tokyo, Bunkyo-ku, Tokyo 113-0033, Japan}
\author{Yong-Chang Lau}
\thanks{Present address: Institute of Physics, Chinese Academy of Sciences, Beijing 100190, China}
\affiliation{Department of Physics, The University of Tokyo, Bunkyo-ku, Tokyo 113-0033, Japan}
\affiliation{National Institute for Materials Science, Tsukuba, Ibaraki 305-0047, Japan}
\author{Vanessa Li Zhang}
\affiliation{School of Physics and Technology, Wuhan University, Wuhan 430072, China}
\author{Goro Shibata}
\affiliation{Department of Physics, The University of Tokyo, Bunkyo-ku, Tokyo 113-0033, Japan}
\affiliation{Materials Sciences Research Center, Japan Atomic Energy Agency, Sayo, Hyogo
679-5148, Japan}
\author{Shoya Sakamoto}
\affiliation{Department of Physics, The University of Tokyo, Bunkyo-ku, Tokyo 113-0033, Japan}
\author{Yosuke Nonaka}
\affiliation{Department of Physics, The University of Tokyo, Bunkyo-ku, Tokyo 113-0033, Japan}
\author{Keisuke Ikeda}
\affiliation{Department of Physics, The University of Tokyo, Bunkyo-ku, Tokyo 113-0033, Japan}
\author{Yuxuan Wan}
\affiliation{Department of Physics, The University of Tokyo, Bunkyo-ku, Tokyo 113-0033, Japan}
\affiliation{Institute for Solid State Physics, The University of Tokyo, Kashiwa, Chiba 277-8581, Japan}
\author{Masahiro Suzuki}
\affiliation{Department of Physics, The University of Tokyo, Bunkyo-ku, Tokyo 113-0033, Japan}
\author{Masashi Kawaguchi}
\affiliation{Department of Physics, The University of Tokyo, Bunkyo-ku, Tokyo 113-0033, Japan}
\author{Masako Suzuki-Sakamaki}
\affiliation{Institute of Materials Structure Science, High Energy Accelerator Research Organization, Tsukuba, Ibaraki 305-0801, Japan}
\affiliation{Graduate School of Science and Technology, Gunma University, Kiryu, Gunma 376-8515, Japan}
\author{Kenta Amemiya}
\affiliation{Institute of Materials Structure Science, High Energy Accelerator Research Organization, Tsukuba, Ibaraki 305-0801, Japan}
\author{Naomi Kawamura}
\affiliation{Japan Synchrotron Radiation Research Institute (JASRI), Sayo 679-5198, Japan}
\author{Masaichiro Mizumaki}
\affiliation{Japan Synchrotron Radiation Research Institute (JASRI), Sayo 679-5198, Japan}
\author{Motohiro Suzuki}
\affiliation{Japan Synchrotron Radiation Research Institute (JASRI), Sayo 679-5198, Japan}
\author{Hyunsoo Yang}
\affiliation{Department of Electrical and Computer Engineering, National University of Singapore, Singapore 117576, Singapore}
\author{Masamitsu Hayashi}
\affiliation{Department of Physics, The University of Tokyo, Bunkyo-ku, Tokyo 113-0033, Japan}
\affiliation{National Institute for Materials Science, Tsukuba, Ibaraki 305-0047, Japan}
\author{Atsushi Fujimori}
\affiliation{Department of Physics, The University of Tokyo, Bunkyo-ku, Tokyo 113-0033, Japan}
\affiliation{Department of Physics and Center for Quantum Science and Technology, National Tsing Hua University, Hsinchu 30013, Taiwan}

\date{\today}

\begin{abstract}
We have studied the Co layer thickness dependences of magnetocrystalline anisotropy (MCA), Dzyaloshinskii-Moriya interaction (DMI), and orbital moment anisotropy (OMA) in W/Co/Pt trilayers, in order to clarify their correlations with each other. We find that the MCA favors magnetization along the film normal and monotonically increases with decreasing effective magnetic layer thickness ($t_\mathrm{eff}$). The magnitude of the Dzyaloshinskii-Moriya exchange constant ($|D|$) increases with decreasing $t_\mathrm{eff}$ until $t_\mathrm{eff} \sim$1 nm, below which $|D|$ decreases. The MCA and $|D|$ scale with $1/t_\mathrm{eff}$ for $t_\mathrm{eff}$ larger than $\sim$1 nm, indicating an interfacial origin. The increase of MCA with decreasing $t_\mathrm{eff}$ continues below $t_\mathrm{eff}$ $\sim$ 1 nm, but with a slower rate. To clarify the cause of the $t_\mathrm{eff}$ dependences of MCA and DMI, the OMA of Co in W/Co/Pt trilayers is studied using x-ray magnetic circular dichroism (XMCD). We find non-zero OMA when $t_\mathrm{eff}$ is smaller than $\sim$0.8 nm. The OMA increases with decreasing $t_\mathrm{eff}$ more rapidly than what is expected from the MCA, indicating that factors other than OMA contribute to the MCA at small $t_\mathrm{eff}$. The $t_\mathrm{eff}$ dependence of the OMA also suggests that $|D|$ at $t_\mathrm{eff}$ smaller than $\sim$1 nm is not related to the OMA at the interface. We propose that the growth of Co on W results in a strain and/or texture that reduces the interfacial DMI, and, to some extent, MCA at small $t_\mathrm{eff}$.
\end{abstract}


\maketitle

\section{Introduction}
Ultrathin film heterostructures that consist of ferromagnetic metal (FM) layers and non-magnetic heavy metal (HM) layers are attracting great interest as various novel phenomena that originate from the strong spin-orbit coupling in bulk and at interfaces have been discovered. For example, efficient current-induced magnetization reversal \cite{Miron2011Nature} and fast motion of magnetic domain walls \cite{Miron2011NM} have been demonstrated in heterostructures with perpendicular magnetic anisotropy (PMA), which are essential in ultra-high-density magnetic memories. These phenomena are attributed to spin-orbit coupling-induced effects such as spin Hall effect \cite{Dyakonov1971JLU, Hirsch1999PRL, Sinova2015RMP, Liu2011Science}, Rashba-Edelstein effect \cite{Rashba1984JetpLett, Edelstein1990SSC}, and Dzyaloshinskii-Moriya interaction (DMI) \cite{Dzyaloshinskii1957JETP,Moriya1960PR}. Among these effects, strong DMI is especially necessary for racetrack memories because it stabilizes chiral N$\rm{\acute{e}}$el domain walls and skyrmions \cite{Bode2007Nature, Parkin2008Science, Hayashi2008Science, Heide2008PRB, Muhlbauer2009Science, Yu2010Nature, Thiaville2012EPL, Emori2013NM, Fert2013NNano, Tacchi2017PRL, Ma2017PRL}. Therefore, solid understanding of these interfacial phenomena is essential to develop spintronic devices with significant PMA and DMI.

Recently, the microscopic origins of PMA and interfacial DMI have been discussed in relation to the orbital moment anisotropy (OMA) \cite{Bruno1989PRB, Yamamoto2017AIPadv} and the magnetic dipole moment in the FM layer \cite{vanderLaan1998JPCM, Kim2018NC}. As OMA should exist both in the FM and HM elements \cite{Solovyev1995PRB}, it is of high importance to identify what role the OMA (of FM and HM elements) plays in PMA and DMI in FM/HM heterostructures.

Here, we study correlation between the magnetocrystalline anisotropy (MCA), DMI and OMA in W/Co/Pt trilayers. The Co layer thickness dependences of MCA and DMI in W/Co/Pt trilayers are studied using vibrating sample magnetometer (VSM) and Brillouin light scattering spectroscopy (BLS), respectively. As for the OMA, we study the Co layer thickness dependences of the spin and orbital magnetic moments of Co in W/Co/Pt trilayers, where W works as a seed layer, and the proximity-induced magnetization in W and Pt in W/Co and Pt/Co bilayers using x-ray magnetic circular dichroism (XMCD). We find that the MCA, DMI, and OMA of Co show different Co layer thickness dependences in the W/Co/Pt trilayers. The origin of these observations will be discussed.

\section{Experiment}
Co thin films sandwiched by W and Pt, i.e. Sub./3 W/$t_\mathrm{Co}$ Co/1 Pt/1 Ru (the numbers denote the nominal thicknesses in nm) were grown on 10$\times$10 mm$^{2}$ thermally oxidized Si substrates by magnetron sputtering at room temperature in a base pressure better than 5$\times$10$^{-7}$ Pa. The top Ru layer is used to protect the trilayers from oxidation. The Ar pressure and RF power were kept constant (0.5 Pa and 50 W) for the sputtering of all the metallic materials. The Co layer thickness ($t_\mathrm{Co}$) was varied from 0.6 to 1.7 nm (0.6, 0.7, 0.8, 1.0, 1.1, 1.2, 1.3, 1.4, and 1.7 nm) in different samples. The thicknesses were determined by a constant deposition rate, which has been calibrated by x-ray reflectivity and tunneling electron microscopy \cite{Liu2015APL}. The magnetic hysteresis loops (in-plane and out-of-plane) of the samples were measured using a VSM at room temperature. The magnetic field was applied up to 1.6 T during the measurement. 

The magnitude of DM exchange constant ($|D|$) was investigated by BLS, with the same measurement setup in our previous studies \cite{Zhang2015APL,Di2015PRL}. The BLS measurements were carried out in the 180$^{\circ}$ back-scattering geometry, using the 514.5 nm radiation of an argon-ion laser and a six-pass tandem Fabry-Perot interferometer. An in-plane saturation magnetic field $H_0$ was applied perpendicular to the incident plane of light, corresponding to the Damon-Eschbach geometry, to obtain the spin-wave dispersion relation. The schematic of the measurement geometry is shown in Fig.~\ref{fig:BLS}(a). The angle between the incident light and the film normal is defined as $\theta$, resulting in the wave vector $k=4\pi\sin\theta/514.5$.

X-ray absorption spectroscopy (XAS) and XMCD measurements at the Co $L_{3,2}$ edges were performed using soft x rays at the helical undulator beamline BL-16A1 of Photon Factory, High Energy Accelerator Research Organization (KEK-PF). The spectra were measured in the total electron yield (TEY) mode. The measurements were performed at room temperature in a vacuum better than  $5 \times \rm{10^{-7}}$ Pa. The magnitude of the magnetic field was set at 5 T and the field was applied parallel to the incident x rays in all measurements. The XAS and XMCD measurements using hard x rays were conducted at BL39XU of SPring-8. The partial fluorescence yield (PFY) and x-ray polarization switching modes were used. The measurements were performed at atmospheric pressure and at room temperature. A magnetic field of up to 2 T was applied during the measurements. In order to obtain the out-of-plane and in-plane components of the magnetic moments, the magnetic field was applied to the sample along the film normal and $30^\circ$ with respect to the sample surface, referred to as out-of-plane and ``in-plane'' magnetic fields hereafter.  Note that the position of the Pt $L_3$ edge (11.563 keV) is excessively close to that of the W $L_2$ edge (11.544 keV): these two peaks will overlap with each other in the W/Co/Pt trilayer and complicate the data analysis. Thus, two W/Co and Pt/Co bilayers were prepared by magnetron sputtering for measuring the W and Pt $L_{3,2}$-edge XAS and XMCD spectra. The bilayer, i.e. Sub./0.6 W/0.8 Co/1 Ru and Sub./0.6 Pt/0.8 Co/1 Ru, were also grown on 10$\times$10 mm$^{2}$ thermally oxidized Si substrates by magnetron sputtering at room temperature and the 1-nm-thick Ru served as the capping to avoid the bilayer from surface oxidation.
	
\section{Results and Discussion}
\begin{figure}
	\begin{center}
		\includegraphics[width=1\columnwidth]{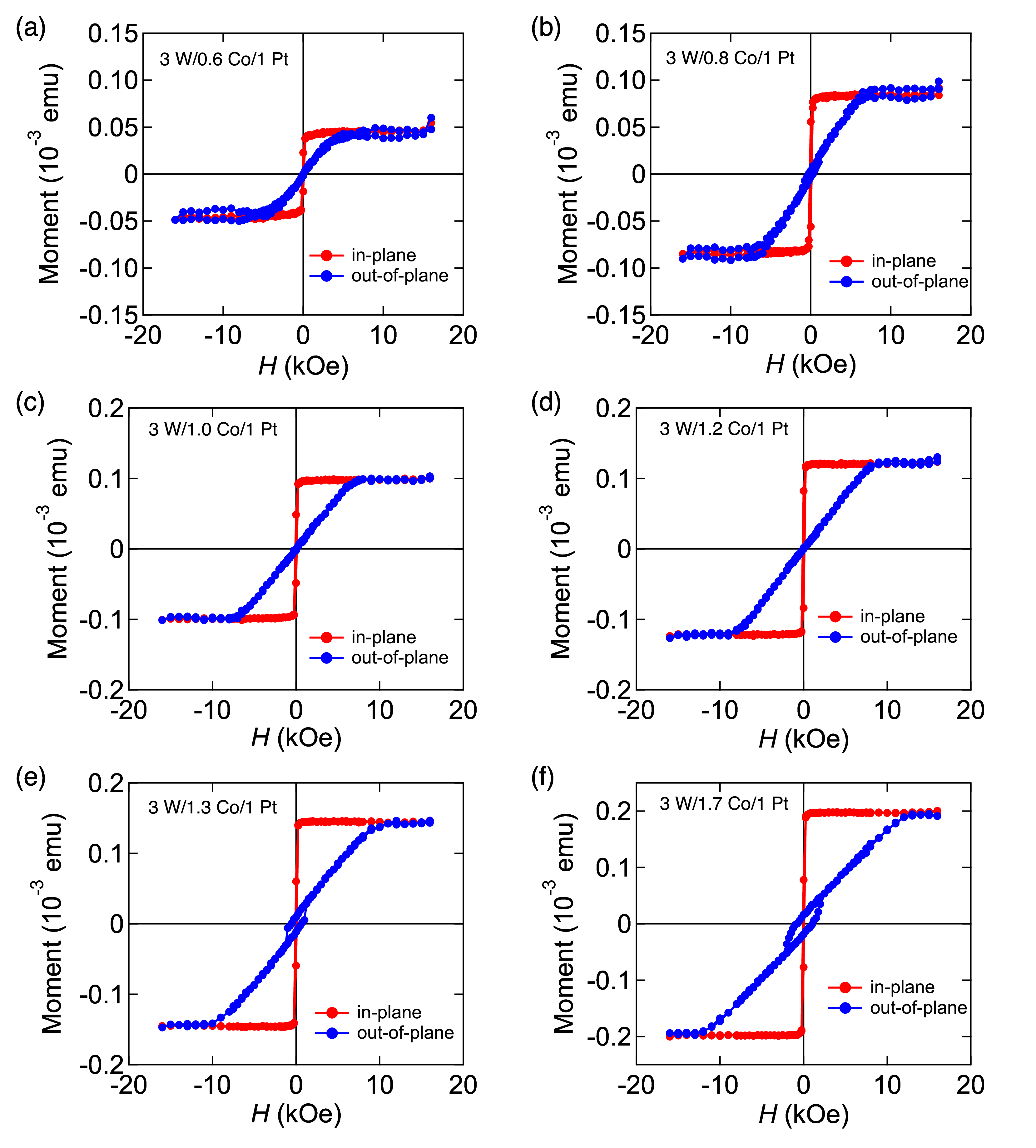}\\
		\caption{Magnetic hysteresis loops of W/Co/Pt trilayers measured by a VSM. The magnetic field is applied perpendicular (blue) and parallel (red) to the film plane. The thickness of the Co layer is (a) 0.6, (b) 0.8, (c) 1.0, (d) 1.2, (e) 1.3, (f) 1.7, respectively.}
		\label{fig:VSM_loops}
	\end{center}
\end{figure}

\begin{figure}
\begin{center}
  \includegraphics[width=1\columnwidth]{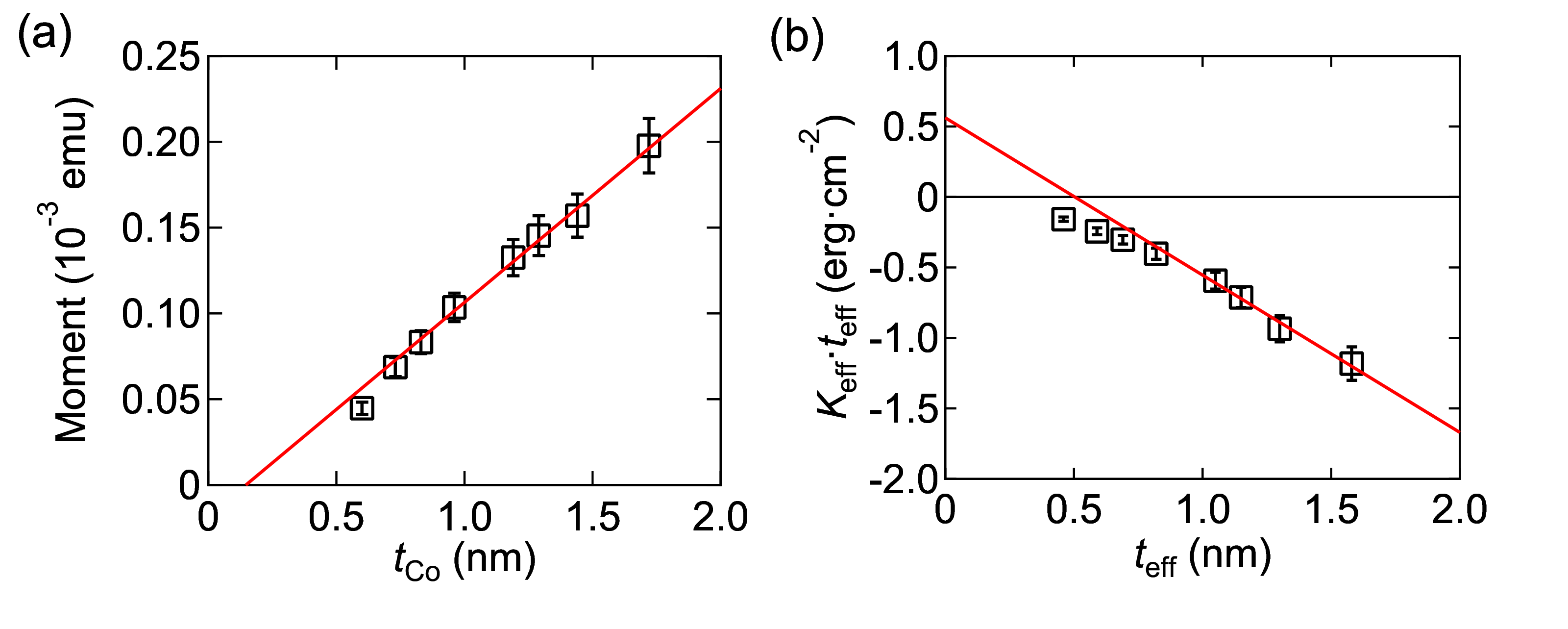}\\
  \caption{(a) Magnetic moment of W/Co/Pt trilayers as a function of Co layer thickness, $t_\mathrm{Co}$. The solid line is a linear fit to the data for $t_\mathrm{Co} > 1$ nm. (b) Product of the effective magnetic anisotropy energy, $K_\mathrm{eff}$, and effective magnetic layer thickness, $t_\mathrm{eff}$, plotted as a function of $t_\mathrm{eff}$. The solid line is a linear fit to the data for $t_\mathrm{eff} > 1$ nm.}
  \label{fig:VSM}
 \end{center}
\end{figure}

Selected magnetic hysteresis loops of the trilayers with different Co thicknesses measured by using the VSM are shown in Fig~\ref{fig:VSM_loops}. The easy axis of the trilayers are confirmed to lie in the film plane. The magnetic moment increases with Co thickness. The magnetic properties of the trilayers determined by the magnetic hysteresis loops are shown in Fig.~\ref{fig:VSM}. The $t_\mathrm{Co}$ dependence of the magnetic moment is shown in Fig.~\ref{fig:VSM}(a). A linear function is fitted to the data with larger weight for films with larger $t_\mathrm{Co}$. The slope of the linear function is proportional to the saturation magnetization (per unit volume) $M_\mathrm{s}$ and the horizontal axis intercept represents the magnetic dead layer thickness $t_\mathrm{D}$. By taking into account the area of the samples, we find $M_\mathrm{s} \sim 1350 \pm 50$ emu$\cdot$cm$^{-3}$ and $t_\mathrm{D} \sim 0.14 \pm 0.05$ nm. The value of $M_\mathrm{s}$ is close to that of bulk Co \cite{Coey_book}. $t_\mathrm{D}$ is typically negative when Co faces a Pt layer due to proximity-induced magnetization \cite{Ueno2015SciRep,Lau2019PRMat}, we infer that a magnetic dead layer at the W/Co interface exists and compensates the negative $t_\mathrm{D}$ at Co/Pt interface. Based on the reported values of proximity-induced moment at Co/Pt interface ($t_\mathrm{D}$ $\sim -0.06$ nm) \cite{Ueno2015SciRep}, the $t_\mathrm{D}$ at W/Co interface is estimated as $\sim 0.20 \pm 0.05$ nm.

The effective magnetic anisotropy energy density, $K_\mathrm{eff}$, is obtained by taking the difference between the integrated areas of the easy-axis and hard-axis magnetization hysteresis loops. With the effective magnetic layer thickness defined by $t_\mathrm{eff} \equiv t_\mathrm{Co} - t_\mathrm{D}$, the product of $K_\mathrm{eff}$ and $t_\mathrm{eff}$, $K_\mathrm{eff} t_\mathrm{eff}$, is given by the following equation \cite{Sahin2013APL,Lau2019PRMat}:
\begin{equation}
\centering
 \label{eq:Keff}
  K_\mathrm{eff} t_\mathrm{eff} = K_\mathrm{I} + (K_\mathrm{B}-2 \pi M_\mathrm{s}^{2})t_\mathrm{eff},
\end{equation}
where $K_\mathrm{B}$ and $K_\mathrm{I}$ represent the bulk and interfacial contributions to $K_\mathrm{eff}$. The $2\pi M_\mathrm{s}^{2}$ term represents the shape anisotropy energy density. $K_\mathrm{eff} t_\mathrm{eff}$ is plotted against $t_\mathrm{eff}$ in Fig.~\ref{fig:VSM}(b). Negative $K_\mathrm{eff} t_\mathrm{eff}$ corresponds to the magnetization easy axis lying along the film plane. A linear function is fitted to the data with larger weight on films with larger $t_\mathrm{eff}$. The slope and the $y$-axis intercept of the linear function represent $K_\mathrm{B}-2 \pi M_\mathrm{s}^{2}$ and $K_\mathrm{I}$, respectively. From the linear fit, we obtain $K_\mathrm{B} \sim (0.8 \pm 0.7) \times 10^{6}$ erg$\cdot$cm$^{-3}$ and $K_\mathrm{I}\sim 0.6 \pm 0.1$ erg$\cdot$cm$^{-2}$. $K_\mathrm{I}$ is smaller than that the values typically reported for structures which include Co/Pt interfaces \cite{Ueno2015SciRep,Lau2019PRMat}. Note that the data show small but systemic deviation from the linear fitting when $t_\mathrm{eff}$ is smaller than $\sim$ 1.0 nm.

\begin{figure}[b]
\begin{center}
  \includegraphics[width=1\columnwidth]{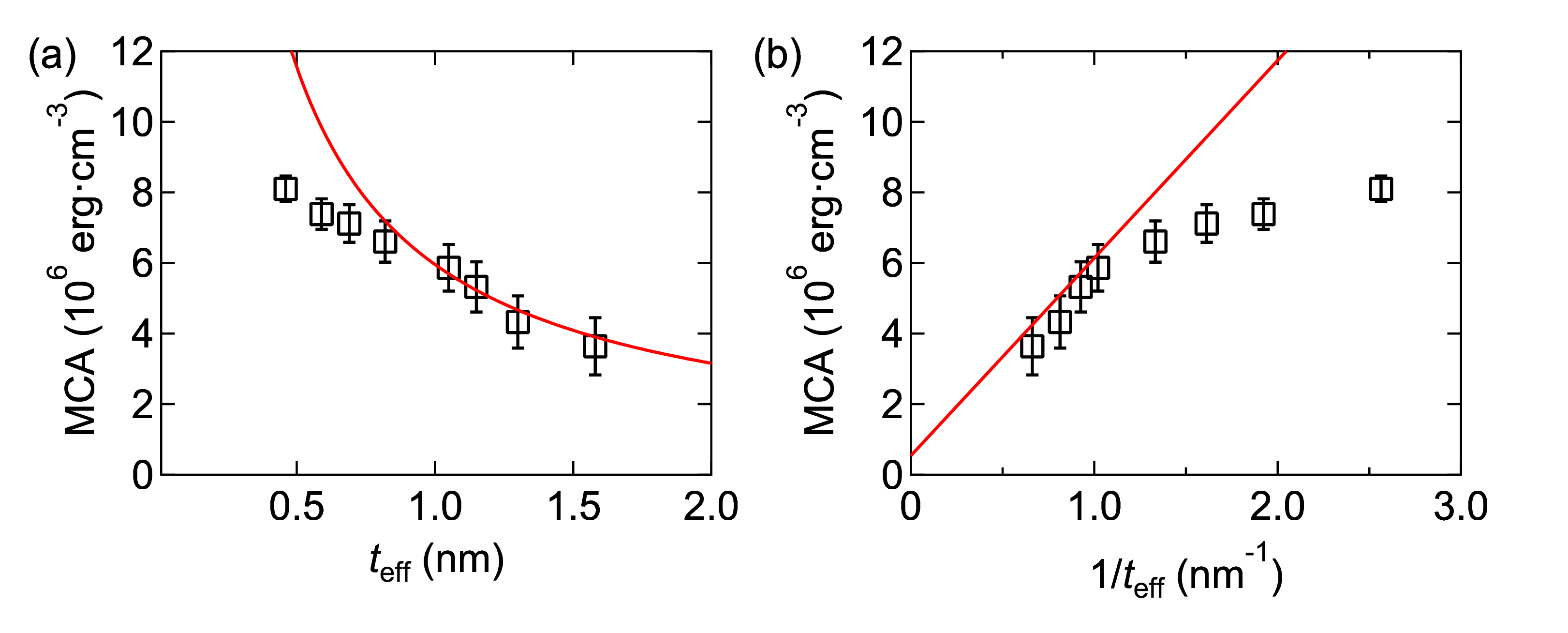}\\
  \caption{(a) $t_\mathrm{eff}$ and (b) $1/t_\mathrm{eff}$ dependence of magnetocrystalline anisotropoy (MCA) in W/Co/Pt trilayers. The solid curves in panels (a) and (b) are calculated using Eq.~(\ref{eq:MCA}) and the values of $K_\mathrm{I}$ and $K_\mathrm{B}$ obtained from the fitting shown in Fig.~\ref{fig:VSM}(b).}
  \label{fig:MCA}
 \end{center}
\end{figure}

The MCA of the trilayers is given by excluding the shape anisotropy from $K_\mathrm{eff}$: 
\begin{equation}
\centering
 \label{eq:MCA}
  \mathrm{MCA} \equiv K_\mathrm{eff} + 2 \pi M_\mathrm{s}^{2} = \frac{K_\mathrm{I}}{t_\mathrm{eff}} + K_\mathrm{B}.
\end{equation}
We plot the $t_\mathrm{eff}$ and $1/t_\mathrm{eff}$ dependences of the MCA in Figs.~\ref{fig:MCA}(a) and (b), respectively. MCA increases monotonically with decreasing $t_\mathrm{eff}$. The calculated MCA using the parameters obtained from the fitting in Fig.~\ref{fig:VSM}(b) is shown by red solid lines in Fig.~\ref{fig:MCA}. Although the MCA is proportional to $1/t_\mathrm{eff}$ for $t_\mathrm{eff}>$ 1 nm, it clearly deviates from the scaling for $t_\mathrm{eff} \lesssim 1$ nm.

\begin{figure}[t]
\begin{center}
  \includegraphics[width=1\columnwidth]{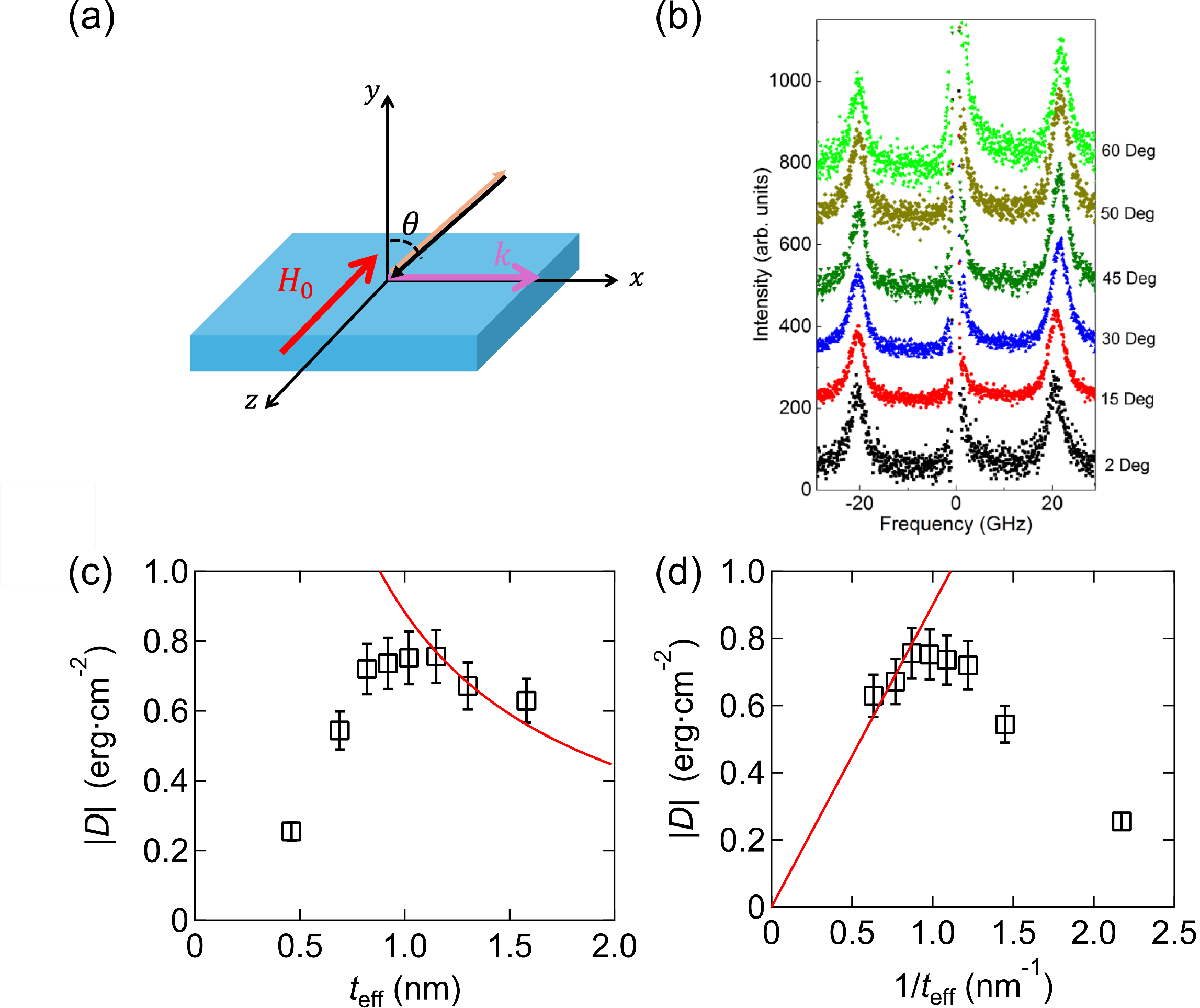}
  \caption{(a) The schematic image of the BLS measurement geometry. The scattering angle is 180$^\circ$. (b) BLS spectra of a 3 W/1.2 Co/1 Pt trilayer measured under different incident light angle $\theta$. (c) $t_\mathrm{eff}$ and (d) $1/t_\mathrm{eff}$ dependence of the magnitude of the Dzyaloshinskii-Moriya exchange constant ($|D|$) in W/Co/Pt trilayers deduced by Brillouin light scattering spectroscopy (BLS) measurements. The solid lines in (a) and (b) show fit to the data from $t_\mathrm{eff}>1$ nm.}
  \label{fig:BLS}
\end{center}
\end{figure}

The properties of DMI in W/Co/Pt trilayers was investigated by BLS measurements. Figure~\ref{fig:BLS}(b) exhibits the BLS spectra for a 3 W/1.2 Co/1 Pt trilayer sample measured with different wavefactor $k$. Due to the lifted chiral degeneracy arising from DMI, the dispersion curves are asymmetrical with respect to $k=0$, where $k$ is the frequency shift. Fitting the measured dispersion data to the equation derived from the Landau-Lifshitz-Gilbert equation \cite{Di2015PRL}:
\begin{widetext}
\begin{equation}
 \label{eq:DMI}
  \omega=\omega_0+\omega_\mathrm{DM}=\mu_0\gamma\sqrt{\left[H_0+Jk^2+\xi\left(kt_\mathrm{eff}\right)M_\mathrm{s}\right]\left[H_0-H_\mathrm{U}+Jk^2+M_\mathrm{s}-\xi\left({kt_\mathrm{eff}}\right)M_\mathrm{s}\right]}-\frac{2\gamma}{M_\mathrm{s}}Dk,
\end{equation}
\end{widetext}
where $\omega_0$ represents the angular frequency in the absence of DMI, and $\omega_\mathrm{DM}$ is the frequency shift induced by DMI. $\mu_0$ is the vacuum permeability, and $\gamma$ is the gyromagnetic ratio set to $194\pm2$ GHz/T (corresponding to a $g$ factor of $\approx$ 2.2). $J=2A/({\mu_0}{M_\mathrm{s}})$, where $A$ is the exchange stiffness constant, and $\xi\left(x\right)=1-\left(1-e^{-|x|}\right)/\left|x\right|$. $H_\mathrm{U}=2\mathrm{MCA}/\left({\mu_0}{M_\mathrm{s}}\right)$ is the uniaxial anisotropic field, with the strength input from the measured MCA values in Fig.~\ref{fig:MCA}. The fitting of the spectra measured with different $k$ (i.e., different $\theta$) yields the magnitude of the Dzyaloshinskii-Moriya exchange constant ($|D|$) based on this equation.

The $t_\mathrm{eff}$ and $1/t_\mathrm{eff}$ dependences of $|D|$ are plotted in Figs.~\ref{fig:BLS}(c) and (d), respectively. $|D|$ increases with decreasing $t_\mathrm{eff}$ until $t_\mathrm{eff} \sim 1$ nm, below which it drops. A similar tendency has  been observed in other HM/FM systems \cite{Cho2015NC, Belmeguenai2018PRB}, which is not in accordance with the simple picture of interface-driven DMI. The $1/t_\mathrm{eff}$ dependence of $|D|$ in Fig.~\ref{fig:BLS}(d) is fitted using a linear function with a larger weight on thicker $t_\mathrm{eff}$, as shown by a red solid line. The parameters obtained by the linear fitting are used to calculate the $t_\mathrm{eff}$ dependence of $|D|$ in Fig.~\ref{fig:BLS}(c), as shown by a red solid line. Figures~\ref{fig:BLS}(c) and (d) display that the experimental data deviates from the linear fitting for $t_\mathrm{eff} < 1$ nm. We note that $|D|$ of W/Co/Pt trilayer grown by molecular beam epitaxy (MBE) has also investigated in a recent study \cite{Jena2021nanoscale}. The value of $|D|$ in a trilayer with 0.7-nm-thick Co layer, $>$ 2 $\mathrm{erg/cm^2}$, is much larger than our samples grown by magnetron sputtering. We suggest the difference can be attributed to the different growth techniques.

\begin{figure}[t]
\begin{center}
  \includegraphics[width=0.8\columnwidth]{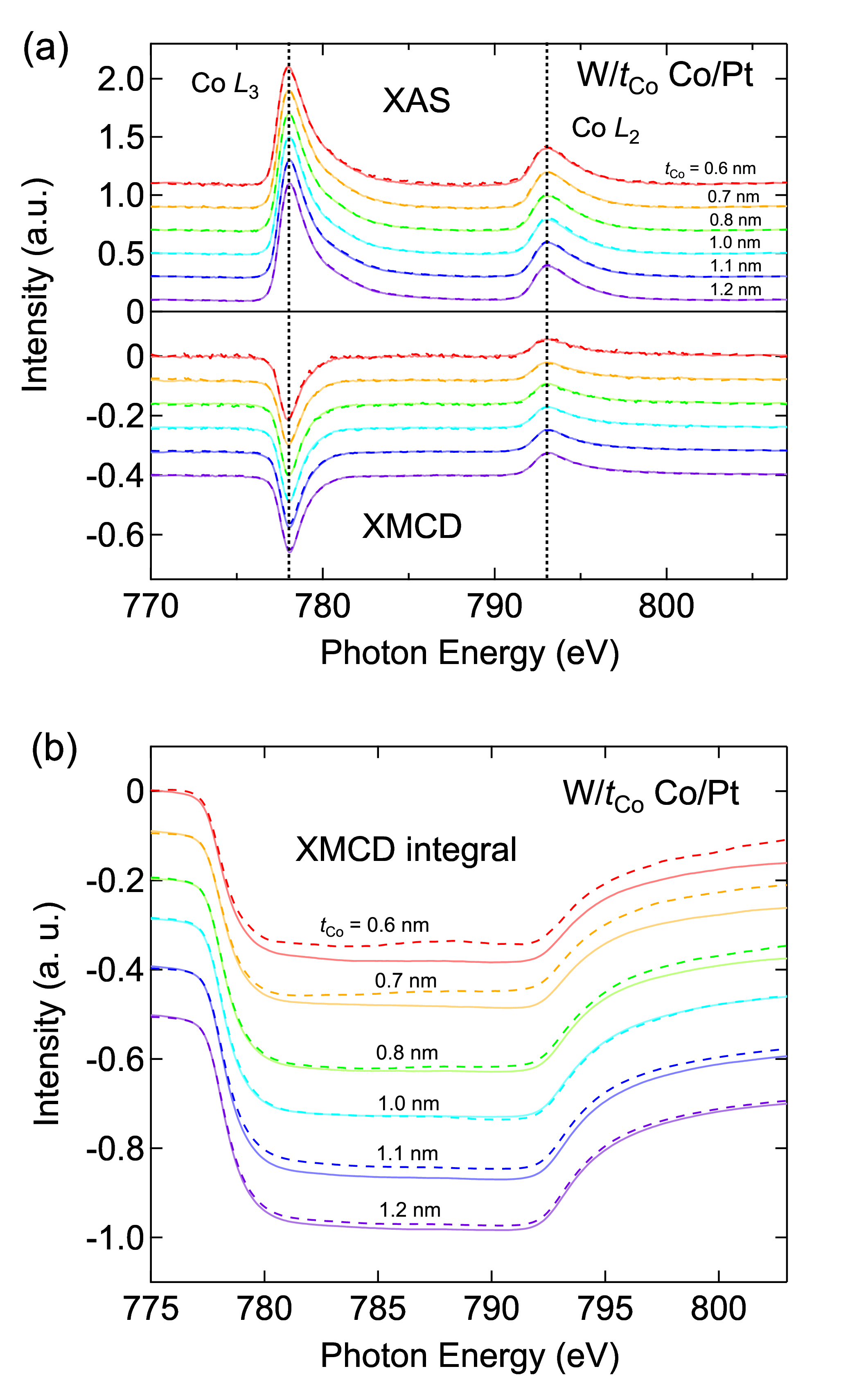}\\
  \caption{XAS, and XMCD spectra of Co in W/Co/Pt trilayers. Spectra after background subtraction are shown. (a) XAS,XMCD and (b) the integrated XMCD spectra at the Co $L_{3,2}$ edges for samples with different Co thicknesses. The spectra obtained under the out-of-plane and ``in-plane'' magnetic fields are plotted by solid and dashed curves, respectively.}
  \label{fig:XMCD}
\end{center}
\end{figure}

\begin{figure*}
	\begin{center}
		\includegraphics[width=1.5\columnwidth]{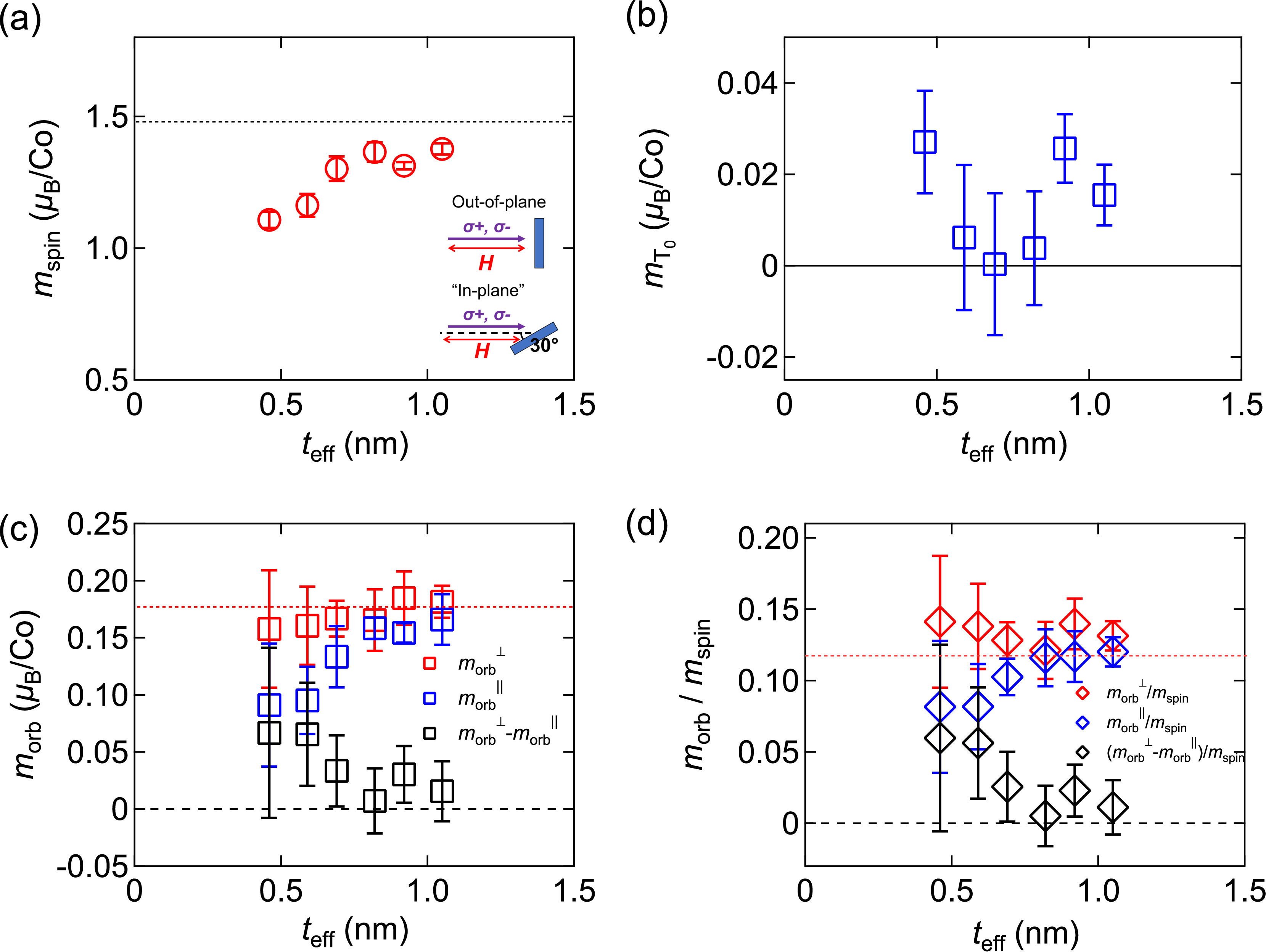}\\
		\caption{$t_\mathrm{eff}$ dependence of (a) the spin moment ($m_\mathrm{spin}$) and (b) the out-of-plane component of magnetic dipole ($m_{\mathrm{T}_0}$). $t_\mathrm{eff}$ dependence of the orbital magnetic moment ($m_\mathrm{orb}$) and normalized orbital magnetic moment ($m_\mathrm{orb}/m_\mathrm{spin}$), for different magnetization directions, are shown in (c) and (d), respectively. The spin, orbital and normalized orbital moment values of bulk hcp Co \cite{Coey_book} are shown using dashed horizontal lines. The schematic images of the XAS and XMCD measurements setup are shown as the inset in panel (a).}
		\label{fig:Moment}
	\end{center}
\end{figure*}

To identify the origin of the $t_\mathrm{eff}$ dependences of MCA and DMI in the W/Co/Pt trilayers, the XAS and XMCD spectra of Co are studied. The setup of the measurements is schematically illustrated in the inset of Fig.~\ref{fig:Moment}(a). Figure~\ref{fig:XMCD}(a) shows the XAS and XMCD spectra at the Co $L_{3,2}$ edges of the W/Co/Pt trilayers measured under a magnetic field of 5 T. The intensity is normalized to the $L_3$ peak after removing a background consisting of two step functions. No obvious peak shift or spectral line-shape change is found in both the XAS and XMCD spectra between different $t_\mathrm{eff}$, suggesting that there is no significant changes in the chemical state of Co, i.e., the oxidation of Co is negligibly small. The solid and dashed curves in Fig.~\ref{fig:XMCD}(a) represent the spectra measured with out-of-plane and ``in-plane'' magnetic fields, respectively. The integrated XMCD spectra, as displayed by the corresponding curves in Fig.~\ref{fig:XMCD}(b), show clear differences between measurements under out-of-plane and ``in-plane'' magnetic field directions. These results indicate that the magnetic moment of Co is anisotropic. 

The effective spin magnetic moment ($m_\mathrm{eff}$) of Co atom is estimated using the XMCD sum rule \cite{Thole1992PRL, Carra1993PRL}:
\begin{equation}
\begin{aligned}
\label{eq:sumrule_spin}
m_\mathrm{eff}&=m_\mathrm{spin}+\dfrac{7}{2}m_\mathrm{T}\\
&=-\dfrac{2\int_{L_{3}}\Delta \mu d\nu-4\int_{L_{2}}\Delta \mu d\nu}{\int_{L_{3,2}} \mu d\nu}n_{h},
\end{aligned}
\end{equation}
where $\Delta \mu$ and $\mu$ are the difference and sum of the XAS spectra obtained using right- and left-handed circularly polarized light, $m_\mathrm{spin}$ is the spin magnetic moment of the Co atom, $m_\mathrm{T}$ is the magnetic dipole, $n_{h}$ is the number of holes in the 3$d$ band of the Co atom. Here, we present magnetic moments in units of Bohr magneton per Co atom using $n_\textrm{h}=2.45$ \cite{Nakajima1998PRL}. The out-of-plane and in-plane components of the magnetic moments are obtained from the integrated XAS/XMCD spectra measured under the out-of-plane and ``in-plane'' magnetic fields, respectively. $m_\mathrm{spin}$ is considered to be isotropic, but $m_\mathrm{T}$ possesses an angular dependence $m_\mathrm{T}=-\dfrac{1}{2}m_{\mathrm{T}_0} \left(1-3\sin^{2}\theta\right)$ \cite{Wu1994PRL,Stohr1999JMMM}. Here, $m_{\mathrm{T}_0}$ represents the out-of-plane component of $m_\mathrm{T}$ and $\theta$ is the angle between the magentization and the film plane.

The estimated values of $m_\mathrm{spin}$ and $m_{\mathrm{T}_0}$ are shown in Figs.~\ref{fig:Moment}(a) and \ref{fig:Moment}(b) as a function of $t_\mathrm{eff}$, respectively. $m_\mathrm{spin}$ deviates from its bulk value, shown by a horizontal dashed line \cite{Coey_book}, and tends to decrease with decreasing $t_\mathrm{eff}$. Such variation of $m_\mathrm{spin}$ with film layer thickness has also been observed in similar systems \cite{Ueno2015SciRep}. $m_\textrm{spin}$ can be fitted against $t_\textrm{Co}$ using the relation $m_\textrm{spin}=\left(1-t_\textrm{D}/t_\textrm{Co}\right)m_\textrm{spin,active}$, where $m_\textrm{spin,active}$ is the active spin magnetic moment, to estimate the $t_\textrm{D}$ in the Co layer. From the fitting, we obtain $t_\mathrm{D} \sim 0.20 \pm 0.03$ nm, which is consistent with the dead layer thickness determined from the VSM measurements. $m_{\mathrm{T}_0}$, which is considerably smaller than $m_\mathrm{spin}$, represents the anisotropic spin-density distribution, and its strength characterizes the anisotropy of the spin-density distribution of the $d$ orbitals. Although it has been reported that $m_{\mathrm{T}_0}$ is related to the emergence of PMA \cite{vanderLaan1998JPCM} and DMI \cite{Kim2018NC}, here its magnitude is considerably smaller than the previous reports \cite{Kim2018NC}.

The orbital magnetic moment ($m_\mathrm{orb}$) of Co atom is estimated using the XMCD sum rule \cite{Thole1992PRL, Carra1993PRL}:
\begin{equation}
\centering
\label{eq:sumrule_orbit}
m_\mathrm{orb}=-\dfrac{4}{3}\dfrac{\int_{L_{3,2}}\Delta \mu d\nu}{\int_{L_{3,2}} \mu d\nu}n_{h}.
\end{equation}
The out-of-plane component of $m_\mathrm{orb}$, $m_\mathrm{orb}^{\perp}$, is estimated from the XAS and XMCD spectra measured under the out-of-plane magnetic field. The in-plane component, $m_\mathrm{orb}^{\parallel}$, is obtained using the spectra measured under the out-of-plane and ``in-plane'' fields according to the relationship,
\begin{equation}
\centering
\label{eq:orbit_angle}
m_\mathrm{orb} (\theta) =m_\mathrm{orb}^{\perp}\sin^{2}\theta+m_\mathrm{orb}^{\parallel}\cos^{2}\theta,
\end{equation}
with $\theta=30^\circ$. The $t_\mathrm{eff}$ dependence of $m_\mathrm{orb}$ is plotted in Fig.~\ref{fig:Moment}(c). Both $m_\mathrm{orb}^{\perp}$ and $m_\mathrm{orb}^{\parallel}$ decrease with decreasing $t_\mathrm{eff}$ from their bulk value. We find the decrease of $m_\mathrm{orb}^{\parallel}$ with $t_\mathrm{eff}$ is stronger than that of $m_\mathrm{orb}^{\perp}$. This difference leads to the OMA of Co which is illustrated by black squares in Fig.~\ref{fig:Moment}(c), showing an increasing trend with decreasing $t_\mathrm{eff}$. The normalized orbital magnetic moment $m_\mathrm{orb}/m_\mathrm{spin}$ is plotted against $t_\mathrm{eff}$ in Fig.~\ref{fig:Moment}(d). The out-of-plane component, $m_\mathrm{orb}^{\perp}/m_\mathrm{spin}$, increases with decreasing $t_\mathrm{eff}$ whereas the in-plane component, $m_\mathrm{orb}^{\parallel}/m_\mathrm{spin}$, decreases. The normalized OMA of Co, illustrated by black diamonds in Fig.~\ref{fig:Moment}(d), increases with decreasing $t_\mathrm{eff}$, especially for $t_\mathrm{eff} \lesssim$ 0.8 nm. These results indicate that the charge redistribution at the HM/Co interfaces takes place and induces OMA. Considering the density of Co crystal as 8.9 g$\cdot$cm$^{-3}$, we estimate the total magnetization of the W/Co/Pt trilayer with $t_\textrm{eff}$=0.9 nm as 1300 emu$\cdot$cm$^{-3}$. This value is in good agreement with that determined by using the VSM.

\begin{figure}
\begin{center}
  \includegraphics[width=1\columnwidth]{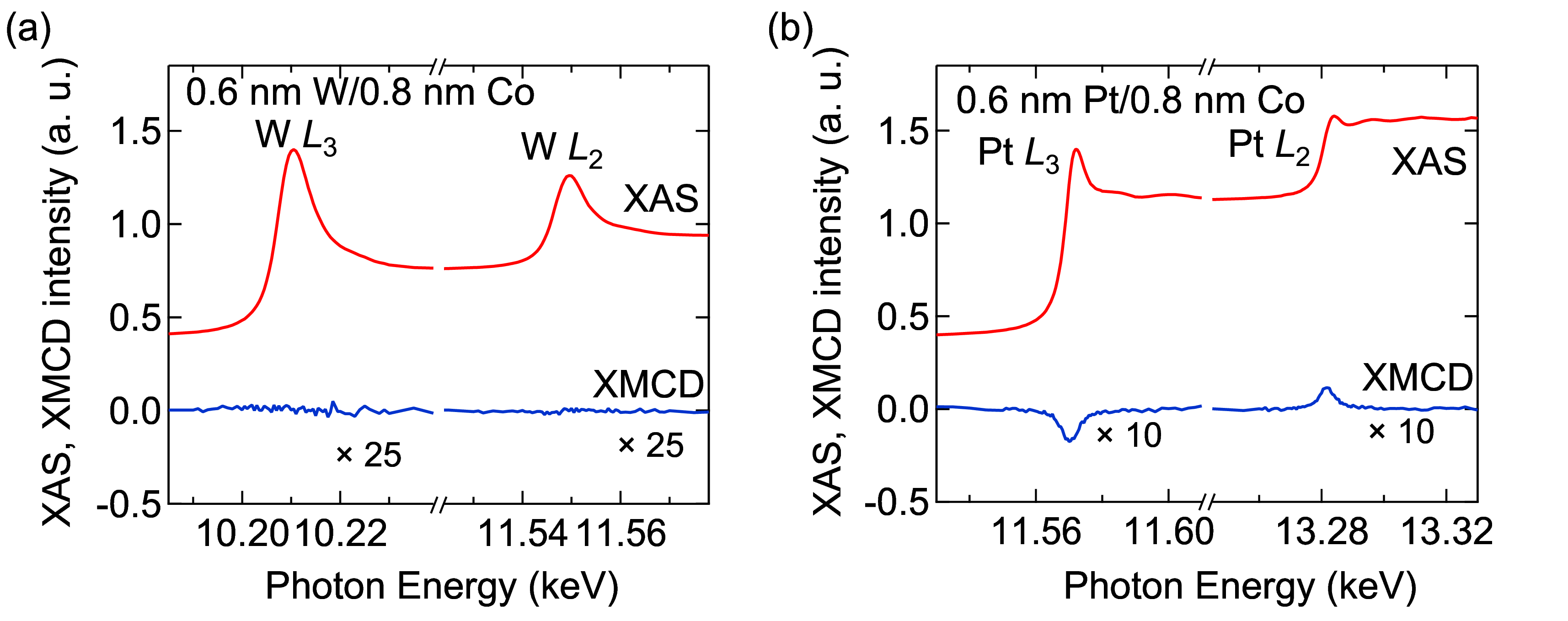}\\
  \caption{XAS and XMCD spectra at the W (a) and Pt (b) $L_{3,2}$ edges in W/Co and Pt/Co bilayers. The intensity of the XMCD spectra of W (Pt) is enlarged by a factor of 25 (10).}
  \label{fig:XMCD_HM}
\end{center}
\end{figure}

We have also performed W and Pt $L_{3,2}$-edge XMCD measurements on W/Co and Pt/Co bilayers to study the proximity-induced magnetization. The XAS and XMCD spectra are shown in Fig.~\ref{fig:XMCD_HM}. XMCD signals are found only in the Pt/Co bilayer. These results indicate that proximity-induced magnetization exists at the Pt/Co interface but does not exist at the W/Co interface.


Now, we discuss the relationship between the MCA, DMI and OMA of Co in W/Co/Pt trilayers. According to Figs.~\ref{fig:MCA} and ~\ref{fig:BLS}, MCA and $|D|$ scale with $1/t_\mathrm{eff}$ for $t_\mathrm{eff} \gtrsim 1$ nm, indicating the interfacial origin of the two properties. Bruno has proposed that the MCA and OMA are proportional to each other in FM monolayers \cite{Bruno1989PRB}. Both MCA and the OMA of Co indeed increase with decreasing $t_\mathrm{eff}$ smaller than $\sim 0.8$ nm. However, the OMA tends to increase more rapidly with decreasing $t_\mathrm{eff}$ than what the Bruno's law predicts from the values of MCA. These results suggest additional contributions to MCA, such as strain or magnetic dipole $m_\mathrm{T}$. Previous studies have indicated that strain in the film texture can weaken the MCA when the magnetic layer thickness is reduced to a few atomic layers \cite{Johnson1996RepProgPhys,Lau2017APL}. The $t_\mathrm{eff}$ dependences of MCA and $K_\mathrm{eff} t_\mathrm{eff}$ are in accordance with such studies. In a very recent study, MCA has been found directly determined by the strain while OMA doesn't follows the prediction by Bruno's rule in a Heusler alloy \cite{Kubota2022PRMater}. These results also suggest that the strain can play a significant role on MCA besides OMA. van der Laan has also shown that $m_\mathrm{T}$ affects MCA in strongly spin-orbit coupled systems, which has been associated with spin-flip virtual excitation \cite{vanderLaan1998JPCM}. Such relationship has been confirmed in recent experiments \cite{Miwa2017NC, Ikeda2017APL, Shibata2018npjQM}. Unfortunately, the XMCD spectra of our W/Co/Pt trilayers does not have sufficient resolution to derive $m_\mathrm{T}$ accurately. 

The DMI, on the other hand, approaches near zero as $t_\mathrm{eff}$ is reduced below $\sim$1 nm. Due to the strong spin-orbit coupling and broken inversion symmetry, the electronic structures in system exhibiting strong DMI are significantly altered by interfacial conditions \cite{Yang2015PRL,Ham2023AdvSci}. Recent studies have shown that DMI is related with $m_\mathrm{T}$ at the HM/FM interfaces \cite{Kim2018NC} and the OMA of the FM atoms \cite{Yamamoto2017AIPadv}. Comparing the results presented in Figs.~\ref{fig:BLS}(c) and \ref{fig:Moment}(c), we consider that such relations do not hold in the current system because the $t_\mathrm{eff}$ dependences of OMA and DMI are opposite to what one expects from the scaling reported in Ref.~\cite{Yamamoto2017AIPadv}. The magnitude of $m_\mathrm{T}$ found in this system is considerably smaller than that reported in Ref.\cite{Kim2018NC} and, therefore, its contribution to DMI, if any, is also likely small. Furthermore, DMI in our sputtered samples is much smaller than that in MBE-grown W/Co/Pt trilayer \cite{Jena2021nanoscale}. Thin films grown by MBE usually show better interfacial roughness and crystal texture compared to films grown by magnetron sputtering. We thus speculate that the DMI is more sensitive to strain effect or the (111) texture of Co. Theoretical studies \cite{Yang2015PRL,Hrabec2014PRB} have indicated that the crystal structure at the HM/FM interface influences the strength of DMI dramatically. In the present case, strain and texture of the Co layer near the W/Co interface may significantly degrade $|D|$ for $t_\mathrm{eff} \lesssim 1$ nm. With further increasing Co thickness, such effects are then mitigated by the Co/Pt interface that favors the (111) texture, resulting in the following decrease of $|D|$. This hypothesis is further supported by another study which also suggests that strain at the Co/Pt interface can significantly modify DMI through charge redistribution within the in-plane $d$-orbitals \cite{Deger2020SciRep}. We note that the same work also points out that the strain-induced tailoring of DMI is related to the charge redistribution within in-plane $d$-orbitals. This mechanism differs from MCA adjustments, which involve altering electron occupation between in-plane and out-of-plane orbitals \cite{Shibata2018npjQM}, indicating the complicated relationship between MCA, DMI, and orbital properties in FM/HM heterostructures.

\section{Summary}

We have studied the effective magnetic layer thickness ($t_\mathrm{eff}$) dependences of magnetocrystalline anisotropy (MCA), Dzyaloshinskii-Moriya interaction (DMI), and orbital moment anisotropy (OMA) in W/Co/Pt trilayers. For $t_\mathrm{eff}$ larger than $\sim$1 nm, MCA and DMI scale with $1/t_\mathrm{eff}$, indicating an interfacial origin. 
However, whereas MCA continues to increase with decreasing $t_\mathrm{eff}$, DMI tends to decrease when $t_\mathrm{eff}$ is reduced below $\sim$1 nm. The OMA of Co deduced from x-ray magnetic circular dichroism (XMCD) measurements is almost zero (below the detection limit) when $t_\mathrm{eff}$ is larger than $\sim$0.8 nm, below which the OMA of Co increases with decreasing $t_\mathrm{eff}$. The rate at which the OMA of Co increases with decreasing $t_\mathrm{eff}$ is larger than what is predicted from the MCA using Bruno's formula. The reduction of DMI with decreasing $t_\mathrm{eff}$ for films with $t_\mathrm{eff} \lesssim 1$ nm, despite the presence of OMA, suggests that other factors contribute to the DMI in this thickness range. We infer that the strain/texture in the Co layer induced by the W underlayer significantly weakens the DMI and, to a lesser extent, the MCA. Further studies are necessary to clarify the latter points. Our results provide a microscopic understanding for designing viable FM/HM-interface-based multifunctional spintronic devices.

\begin{acknowledgements}
 We thank H. Shimazu for samples preparation. This work was supported by Grants-in-Aid for Scientific Research from JSPS (Grant Nos. 15H02109, 15H05702, 16H03853, 20K14416, and 22K03535) and by the National Science and Technology Council of Taiwan under a Grant No. 113-2112-M-007-033. The XMCD experiment was performed at BL-16A of KEK-PF with the approval of the Photon Factory Program Advisory Committee (proposal Nos. 2016S2-005 and 2016G066) and at BL39XU of SPring-8 with the approval of the Japan Synchrotron Radiation Research Institute (JASRI) (proposal Nos. 2017A1048 and 2018A1058). M.H. and A.F. are adjunct members of Center for Spintronics Research Network (CSRN), the University of Tokyo, under Spintronics Research Network of Japan (Spin-RNJ). Z.C. is supported by Materials Education program for the future leaders in Research, Industry, and Technology (MERIT) and JSR Fellowship, The University of Tokyo. Y.-C.L. is supported by JSPS International Fellowship for Research in Japan (Grant No. JP17F17064). S.S. and Y.-X.W. acknowledges financial support from Advanced Leading Graduate Course for Photon Science (ALPS). S.S. acknowledges financial support from the JSPS Research Fellowship for Young Scientists. A.F. acknowledges the support from the Yushan Fellow Program under the Ministry of Education of Taiwan.
\end{acknowledgements}

\bibliography{Manuscript_W_Co_Pt_XMCD}

\clearpage

\end{document}